\documentclass[twocolumn,10pt,letterpaper]{article}
\usepackage[top=1.9cm, bottom=1.9cm, left=1.7cm, right=1.7cm]{geometry}
\usepackage{booktabs} %
\usepackage{color}
\usepackage{times}
\usepackage{subfigure}
\usepackage{paralist}
\usepackage[numbers,sort&compress]{natbib}
\usepackage[keeplastbox]{flushend}  %
\usepackage{graphicx}  %
\usepackage[numbers,sort&compress]{natbib}

\usepackage{courier}  %
\usepackage[hyphens]{url}  %
\usepackage{graphicx}  %
\graphicspath{ {images/} }
\usepackage{array}
\usepackage{makecell}
\usepackage{multirow, booktabs}
\usepackage{siunitx}
\usepackage{ifthen} 
\usepackage[bf]{caption} 

\usepackage[hang,flushmargin]{footmisc}

\usepackage{ifthen}
\newif\ifshort
\shortfalse %
\ifshort
	\newcommand{\isShort}{true}
\else
	\newcommand{\isShort}{false}
\fi
\newcommand{\shortVer}[1]{\ifthenelse{\equal{\isShort}{true}}{{#1}}{}}
\newcommand{\longVer}[1]{\ifthenelse{\equal{\isShort}{false}}{{#1}}{}}

\usepackage[english]{babel}

\newcommand{\descr}[1]{\smallskip\noindent\textbf{#1}}

 \makeatletter
 \def\url@leostyle{%
   \@ifundefined{selectfont}{\def\UrlFont{\small}}%
   {\def\UrlFont{}}%
 }
 \makeatother
 \renewcommand{\footnoterule}{%
   \kern -3pt
   \hrule width 1in 
   \kern 2pt
 }
 
 \let\OLDthebibliography\thebibliography
\renewcommand\thebibliography[1]{
  \OLDthebibliography{#1}
  \setlength{\parskip}{0.5pt}
  \setlength{\itemsep}{0pt plus 0.3ex}
}

\usepackage{titlesec}
\titlespacing*{\section}{0pt}{*2}{5pt} 
\titlespacing{\subsection}{0pt}{*2}{4pt}
\titlespacing{\subsubsection}{0pt}{*1.5}{3pt}

\usepackage{etoolbox} 
\makeatletter
\patchcmd\maketitle{\@makefntext}{\@@@ddt}{}{}
\patchcmd\maketitle{\rlap}{\mbox}{}{}
\makeatother

\definecolor{darkgreen}{RGB}{135,0,40}
\usepackage[bookmarks=false, colorlinks=true, plainpages = false,  linkcolor = darkgreen, citecolor = darkgreen, urlcolor = darkgreen, filecolor = black]{hyperref}

\begin{document}
\title{\bf Of Wines and Reviews: Measuring and Modeling the\\Vivino Wine Social Network\thanks{A preliminary version of this paper appears in the Proceedings of the IEEE/ACM International Conference on Advances in Social Networks Analysis and Mining (ASONAM 2018). This is the full version.}}

\author{
Neema Kotonya$^1$, Paolo De Cristofaro$^2$, and Emiliano De Cristofaro$^1$\\
{\normalsize $^1$University College London $\;\;$ $^2$Tipicamente Wine Blog}\\
{\normalsize neema.kotonya.13@ucl.ac.uk, e.decristofaro@ucl.ac.uk, paolodecri@alice.it}
}
\date{}

\thispagestyle{plain}
\pagestyle{plain}

\maketitle

\begin{abstract}
This paper presents an analysis of social experiences around wine consumption through the lens of Vivino, a social network for wine enthusiasts with over 26 million users worldwide. We compare users' perceptions of various wine types and regional styles across both New and Old World wines, examining them across price ranges, vintages, regions, varietals, and blends. Among other things, we find that ratings provided by Vivino users are not biased by cost.
We then study how wine characteristics, language in wine reviews, and the distribution of wine ratings can be combined to develop prediction models. More specifically, we model user behavior to develop a regression model for predicting wine ratings, and a classifier for determining user review preferences. %
\end{abstract}

\section{Introduction}

Over the years, research has shown that food and alcohol consumptions are strongly shaped by social influences~\cite{higgs2016social,tiu2001food}, often mirroring those of people with shared social connections \cite{silva2014you}. 
In fact, eating and drinking habits are increasingly part of our social media footprints.
Moreover, regional factors, such as economic, cultural, and lifestyle variations, inevitably affect perceptions and choices~\cite{atkins2016food}. 
Therefore, the rise of dedicated social apps, as well as the growing use of social media to discuss and share such habits, offer new opportunities to elicit valuable insight at a large scale.

Previous work has studied the cultural and socio-economic factors determining what we eat and drink~\cite{eufic}. 
More recently, researchers have analyzed food consumption on mainstream social networks such as Twitter \cite{abbar2015you} and Instagram \cite{mejova2016fetishizing}, as well as beer habits on social networks like Untappd~\cite{chorley2016pub}. 
However, to the best of our knowledge, wine consumption has long been overlooked, even though it has played an important societal role in many cultures for thousands of years~\cite{trevisan2011wine}.

Aiming to bridge this gap, this paper studies Vivino, a social network application for wine enthusiasts with over 26 million active users worldwide.
Vivino provides a platform for reviewing and rating wines, forums for sharing experiences and knowledge, as well as a marketplace for buying and selling wines. We focus on Vivino as it allows us to perform a large-scale measurement of wine consumption as well as social network analysis geared to building novel regression and classification models.

Specifically, we focus on two main research objectives: (1)  Characterizing the Vivino social network in terms of how regional factors, user biases, and wine characteristics affect wine ratings; and (2) Integrating wine characteristics, reviews, and wineries on Vivino to develop practical models for predicting wine ratings and user reviews.

We crawl Vivino to collect data about wines, including prices and ratings, as well as attributes, such as their vintage, type, regional style, and the winery from which they originate. We also collect user data, including ratings,  reviews, taste profile, ranking, country, followers, and the users whom they follow. 
Next, we study the relationship between ratings and a number of wine characteristics. We examine users' biases as to which wines they drink and how they score them, as well as the language used in both the user biographies and the wine reviews. 
Lastly, we integrate the findings of our analysis to develop systems for predicting users' ratings and reviews.
In other words, we show how to use the patterns uncovered in the data analysis of Vivino wines, users and user-generated content, to generate: (i) a predictive model for wine ratings, and (ii) a model for categorizing user preferences. 

\descr{Main findings.} Overall, we find that: 
\begin{enumerate}
    \item The ratings and the reviews supplied by Vivino users display the same rich knowledge of wines as professional wine reviews. However, unlike the latter, Vivino users' ratings do not seem to be heavily affected by wine prices; 
    \item Vivino users have an affinity for rating local wines. There are also strong geographical similarities in how wines from adjacent countries are rated;
    \item Vivino's user-generated data produced accurate, practical models for the prediction of wines. The regression models for predicting wine ratings achieved $R^2$ scores $>$ 0.60, and the user preference classifier has a mean accuracy score of almost 80\%. 
\end{enumerate}

\section{Related Work}
Previous work has studied online diaries of food and drink consumption habits,
e.g., examining how dietary choices are linked to food-related tweets by Twitter users in the 
United States. 
Abbar et al.~\cite{abbar2015you} find a strong correlation between food mentions on Twitter with obesity and diabetes statistics,
while Mejova et al.~\cite{mejova2015foodporn} present a large-scale analysis of pictures taken at US restaurants.
Sharma and De Choudhury~\cite{sharma2015measuring} investigate how nutritional information about food is communicated through Instagram.

Researchers have also analyzed recreational eating on social networks. 
Guidry et al.~\cite{guidry2015mcdonaldsfail} study how social media hashtags are used to express approval and disapproval of fast food chains, while Mejova et al.~\cite{mejova2016fetishizing} collect a dataset of Instagram conversations including certain hashtags, and characterize three kinds of users: singletons, residents, and travelers, %
showing that travel drives emotions associated by users to hashtags.

Another line of work introduces models for predicting the cuisine and the geographical origin of a dish from its 
recipe~\cite{sajadmanesh2016kissing,ghewaripredicting}, collecting data from wiki-like pages, rather than social networks. 
These mostly aim to develop supervised learning models to predict the cuisine, via multi-class learning.
\longVer{This is similar to one of the models we propose, although we also use a regression based one to determine wine ratings in real-value terms. 
Moreover, the datasets they use are significantly smaller than ours, i.e., thousands vs millions of entries.
We also find that confusion matrices presented in previous work show that the most commonly misclassified cuisines are from the same regions of the world, thus, the presumable source of their ingredients is from the same locations. In our work, we aim to discover if there was any detectable confusion between Old and New World wines; besides looking at prediction models, we also consider models for recommendation. These allow us to learn more about users, as they introduce users to new genres, in our case, wine types, which they might have not previously considered.

} Overall, while numerous studies focus on eating habits and food consumption, fewer analyze the behavior driving consumption of drinks, in particular alcoholic beverages. A notable exception is the work by Chorley et al.~\cite{chorley2016pub}, who analyze the beer-oriented app Untappd. %
Specifically, they study beer ratings and highlight that Untappd users express a generally favorable opinion of  lagers and ales. They also find correlation between scores assigned to beers by American and European drinkers,
showing that Untappd displays a power-law distribution like most real networks~\cite{faloutsos1999power}.

\section{Datasets}\label{sec:datasets}

\subsection{Background: Vivino}\label{subsec:vivino}

Vivino is a wine marketplace and online community for wine enthusiasts, which is available both as a web and a mobile application. It was founded in 2009 by Heini Zachariassen, with his colleague Theis Sondergaard joining the venture in 2010. Vivino has grown rapidly since then, boasting 29 million users as of March 2018. In a nutshell, the application allows users to review and purchase wines through third-party vendors. 
The mobile application also provides a {\em wine scanner} functionality, i.e., users can upload pictures of wine labels and access reviews and details about the wine and the winery from which it originates.

Vivino is really a social network, as it allows wine enthusiasts to communicate with and follow each other, as well as share reviews and recommendations. As of March 2018, Vivino reports featuring 9.2M wines (including dessert and port wines), covering a multitude of wine styles, grapes, and geographical regions, as well as 89.4M ratings and almost 29.9M reviews.\footnote{\url{https://www.vivino.com/about}}
Users can also earn a variety of rewards for their activity, e.g., receiving {\em likes} on their posts, and get prompted to unlock achievements, e.g., if they ``scan a wine from Argentina.'' High-performing reviewers also become {\em ambassadors} or receive labels like  ``Top Ten in Country Californian Meritage.''\longVer{

} In terms of reviews, Vivino claims that their 5-star rating system (with 0.5 granularity) has a good correlation with Robert Parker's 100 point scale \cite{parker2002parker}, and argues that its users are able to produce a greater number of ratings than the seven most prolific wine experts (in fact, Vivino users have produced ratings for 1.4M wines in the period between 2011 and 2015, while only 370k wines received expert ratings).\footnote{\url{https://www.vivino.com/wine-news/vivino-5-star-rating-system}}

\subsection{Crawler}
Vivino does not offer an API to collect data from their site, therefore, we gathered data about wines, wineries, and users from the Vivino website using a custom web crawler in Python, relying on the {\em Selenium} and {\em requests} packages. %
To avoid generating an extensive amount of traffic, causing possible issues for the site operators, we throttled the crawler to 0.2 requests per second, and ran it over five months (November 2016 to March 2017).

\begin{table*}[t]
	\setlength{\tabcolsep}{8pt}
    \centering
    \small
    \begin{tabular}{ l rr rr rr rr }
    \toprule
    \textbf{Country } &
    \multicolumn{2}{c}{\textbf{ Wines }} &
    \multicolumn{2}{c}{\textbf{ Users }} &
    \multicolumn{2}{c}{\textbf{ Wine Reviews }} &
    \multicolumn{2}{c}{\textbf{ Wineries }}\\
    \midrule
    Argentina &		43.5k & 4\%		& 8.8k & 6\%		& 35.5k & 5\%	& 0.19k & 2\%\\
    Australia & 		56.0k &5\%		& 9.5k & 6\% 	& 55.0k &  7\% 	& 0.41k & 5\%\\
    Brazil & 		3.9k &0\%		& 8.6k & 6\% 	& 0.8k & 0\% 	& 0.01k & 0\%\\
    Canada & 		7.9k &1\%		& 9.4k & 6\% 	& 9.5k & 1\% 	& 0.05k & 1\%\\ 
	China &			0.0k &0\%		& 9.9k & 7\%	 	& 0.0k & 0\% 	& 0.00k & 0\%\\   
    France & 		252.0k &24\% 	& 9.9k & 7\%  	& 139.6k & 18\% 	& 1.89k & 23\%\\
	Germany & 		23.6k &2\%		& 9.7k & 7\%  	& 8.4k & 1\% 	& 0.08k & 1\%\\
    Italy & 			222.7k &21\% 	& 10.7k & 7\% 	& 132.1k & 17\% 	& 1.76k & 21\%\\
    Portugal &		52.1k &5\%		& 9.4k & 6\% 	& 20.4k & 3\% 	& 0.44k & 5\%\\
    Romania	&		2.1k &0\%		& 9.5k & 6\% 	& 0.5k & 0\%		& 0.02k & 0\%\\
    Russia &			0.4k	 &0\% 		& 10.8k & 7\% 	& 0.0k & 0\%		& 0.00k & 0\%\\
    South Africa &	41.2k &4\%		& 9.8k & 7\% 	& 50.3k & 7\% 	& 0.31k & 4\%\\
    Spain &			73.6k &7\%		& 10.5k & 7\%	& 48.1k & 6\% 	& 0.62k & 8\%\\
    UK & 			1.3k &0\%		& 10.9k & 7\% 	& 1.2k & 0\% 	& 0.00k & 0\%\\
    USA &			173.0k &16\%		& 9.8k  & 7\%	& 209.8k & 27\% 	& 1.70k & 21\%\\
    Others 		&	109.4k &10\%		& 6.7k & 5\% 	& 60.7k & 8\%	& 0.76k & 9\%\\
    \midrule
    {\em Total} 			& 1.06M &		& 147.1k &&  771.9k 	&&  8.3k  \\
    \bottomrule
    \end{tabular}
    \caption{Summary of the wine, user, wine reviews (initial posts, not replies) and winery datasets collected by crawling the Vivino website, broken down by country. }
    \label{tab:main_stats}
    \vspace*{-0.1cm}    
\end{table*}

\subsection{Wines, Users, Reviews, and Wineries}
Our crawl yields four datasets, respectively, containing wines, users, reviews, and wineries, as discussed next.

\descr{Wines dataset.} In total, we collected data for 1.06M wines. We gathered the following attributes:
{\em Name} (of the wine), 
{\em Type} (i.e., white, red, ros\'e, port, dessert, or sparkling),
{\em Vintage} (i.e., year of production),
{\em Average Price},
{\em Average Rating},
{\em Number of Ratings},
{\em Ratings Breakdown} (i.e., the number of 1.0, 2.0, 3.0, 4.0, and 5.0 stars received),
{\em Country},
{\em Region} (e.g., Bourgogne),
{\em Regional Style}, (e.g., Spanish Rioja)
{\em Winery},
{\em Food Pairings} (e.g., lamb), and
{\em Grapes} (e.g., merlot).

Note that some fields are sparsely populated, e.g., {\em Regional Style}. Also, the {\em Average Price} really depends on the availability at suppliers working with Vivino, and 474.3k wines are not listed at any supplier, thus the corresponding field is empty.

\descr{Users dataset.} Next, we selected the 10k most active users from each of the 15 countries with the highest level of wine consumption  according to the International Organization of Vine and Wine (OIV) \cite{oiv2016}, i.e., Argentina, Australia, Brazil, Canada, China, France, Germany, Italy, Portugal, Romania, Russia, South Africa, Spain, UK, and USA. %
We did so since the majority of Vivino users do not submit any review, we wanted to ensure that we capture a non-negligible number of reviews per user. 
In order to get the top users, we used Vivino's ranking system, which ranks users by country according to their activity (i.e., number of reviews) and contributions to the platform (i.e., posts receiving likes and comments).
Out of the 150k top users (10k for 15 countries), we acquired data for 147.1k (98\%). 
We were unable to collect all the top users because of web crawling restrictions and time limitations. The rest was missed due to crawler failure, errors, and/or time limitations. 

Overall, for each user, we collected: {\em Username}, {\em Biography}, {\em Country}, {\em Ranking} (i.e., how the user ranks in terms of rating/review contributions compared with other users in their country), {\em Number of Followers} the user has acquired, {\em Number of Users Followed}, {\em Taste Profile} (the regional styles reviewed by the user, with counts and average ratings), the {\em Total Number of Ratings} supplied by the user to Vivino, and whether the user has been {\em Featured} by Vivino -- a special status where the user's profile is promoted on the application to other users. 

\descr{Reviews dataset.} We also collected the reviews posted by each of the top users in our user dataset, gathering 771.9k review posts for 86.6k unique wines. (In other words, we created a database with wine reviews collected using the IDs of a sample of the wines scraped.)
For each review, we gathered: {\em Wine}, {\em Vintage}, {\em Content} (which includes the author's username and the number of ratings provided by them), {\em Date}, and {\em Replies}. %
Overall, we gathered 771.9k reviews, which were posted by 370k unique authors. Additionally, these initial review posts garnered 617.7k replies collectively.

\descr{Wineries dataset.} Similar to the wines dataset, we also got records for 8.3k wineries, which represent the 1.06M wines scraped from Vivino. %
Specifically, we retrieved: 
{\em Basic Details} (i.e., name, number of wines produced, URL of its Vivino profile page),
{\em Ratings} (number and average score),
{\em Wine Maker} (if any),
{\em Location} (GPS, region, and country), and
{\em Websites} (Twitter, Facebook, and official websites).

\subsection{General Characterization} 
In Table~\ref{tab:main_stats}, we present a summary of our datasets.
As discussed above, we have data for 1.06M wines 
as well as 147.1k user profiles,
along with 771.9k reviews and 8.3k profiles of wineries. %

\descr{Countries.} The entire wine dataset includes wines from 49 distinct countries. Interestingly, 98\% of them (1.04M) are produced in 21 different countries\footnote{As per most wines: France, Italy, USA, Spain, Australia, Portugal, Chile, Argentina,
South Africa, Germany, New Zealand, Austria, Canada, Hungary, Israel, Brazil,
Greece, Romania, Georgia, Uruguay, Mexico, Switzerland.}
The top four countries, i.e., France, Italy, Spain. and the United States, account for more than two third of all the wines on Vivino (69.1\%). This is not surprising since these are also the four largest wine producing countries, accounting for an estimated 59\% of the worldwide wine production in 2016~\cite{oiv2017}.

\begin{figure*}[t]
 \centering
    \subfigure[By Country]{\includegraphics[width=0.375\textwidth]{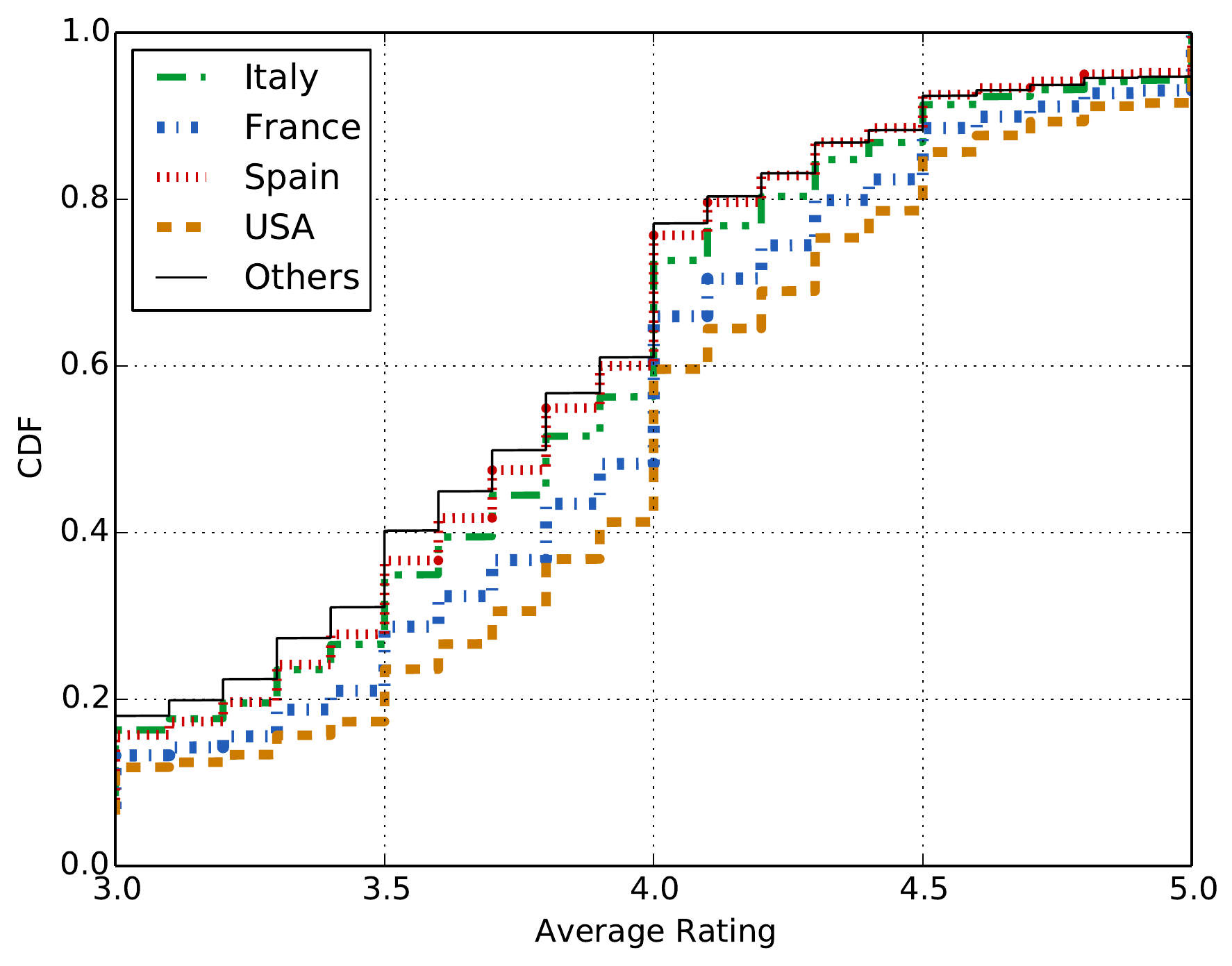}}
    \subfigure[By Rating]{\includegraphics[width=0.375\textwidth]{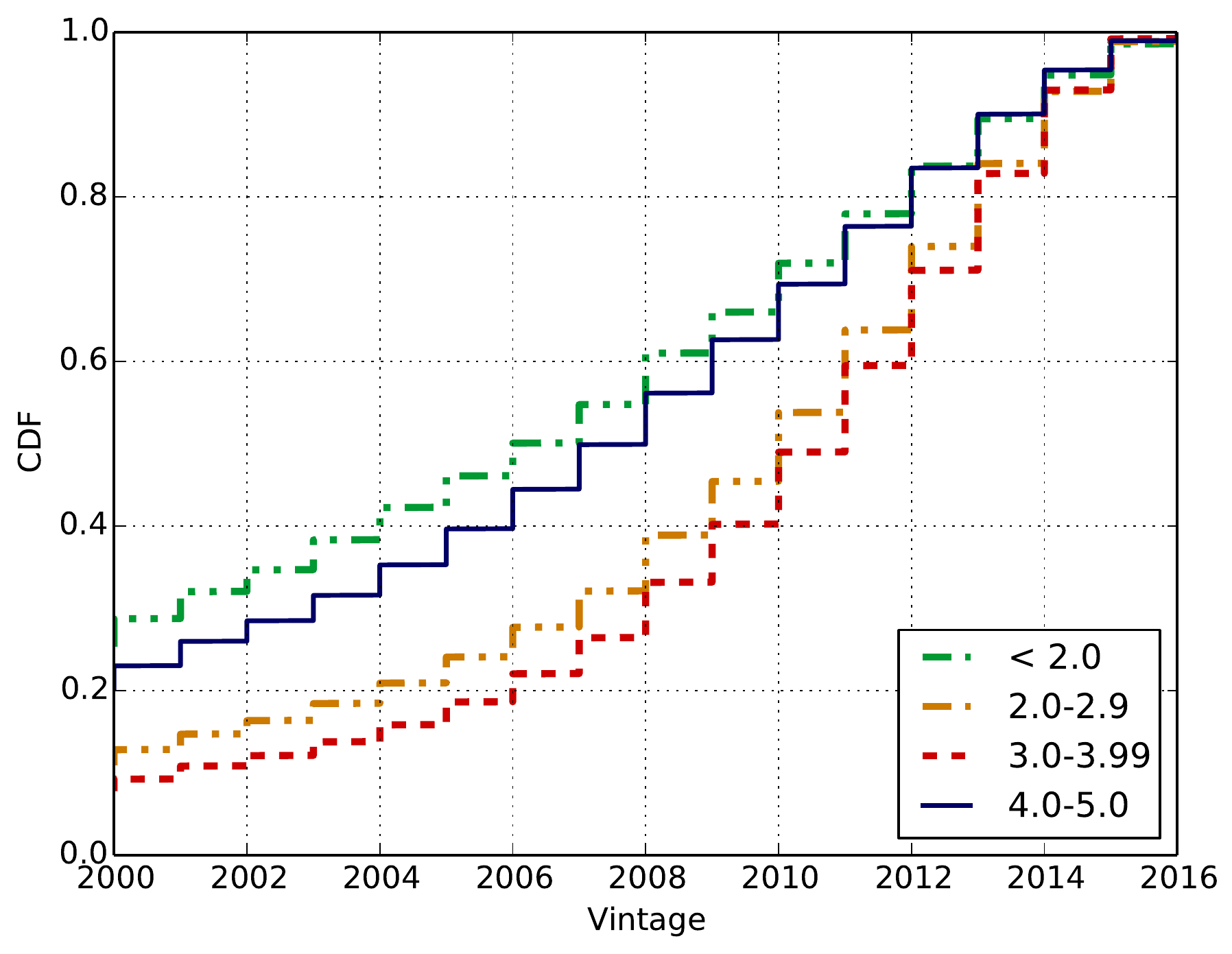}}
    \caption{CDFs of average ratings by country and vintages by rating.}
    \label{fig:ratings_cdf}
\end{figure*}

\begin{figure*}[t]
\centering
    \subfigure[By Country]{\includegraphics[width=0.375\textwidth]{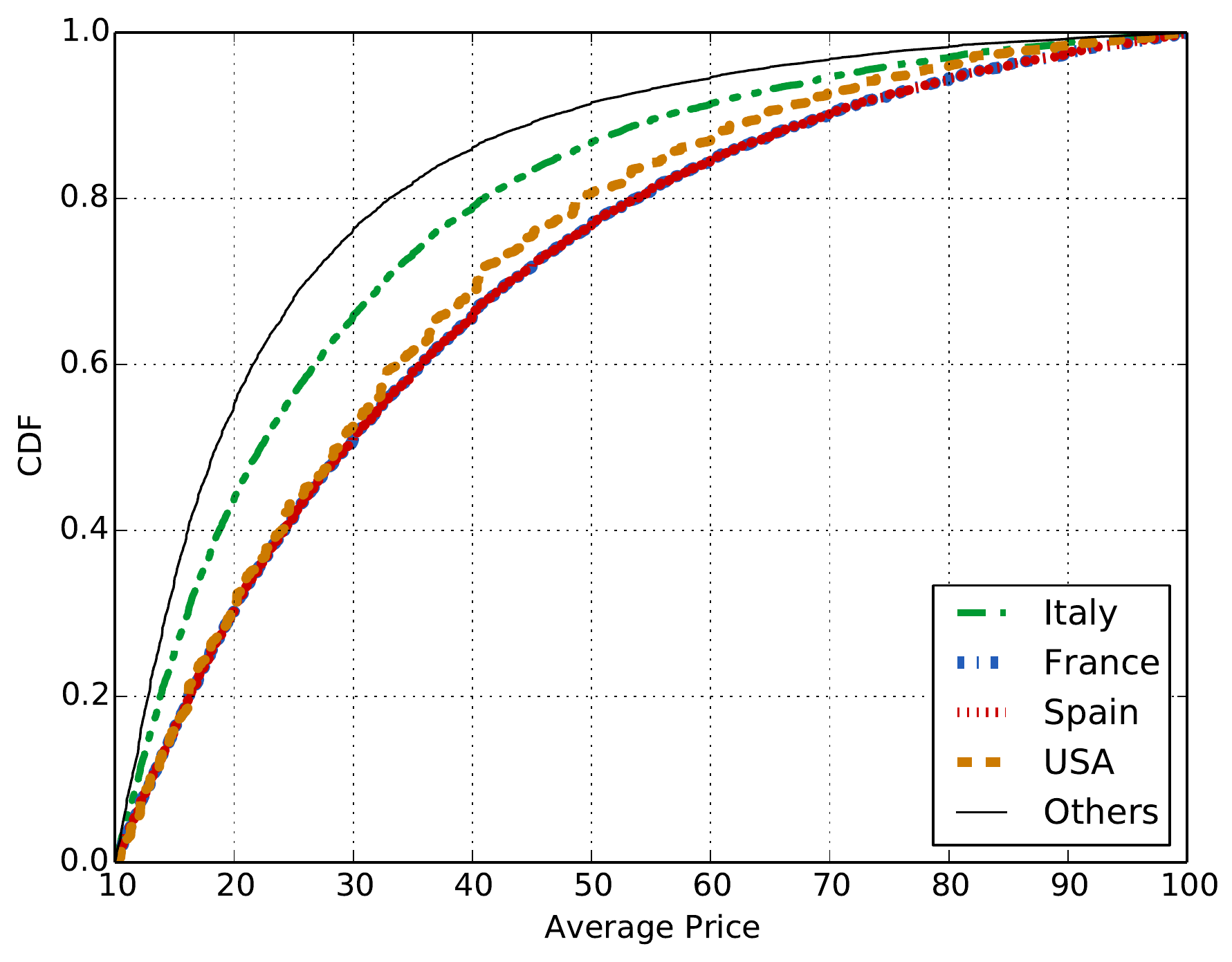}}
     \subfigure[By Rating]{\includegraphics[width=0.375\textwidth]{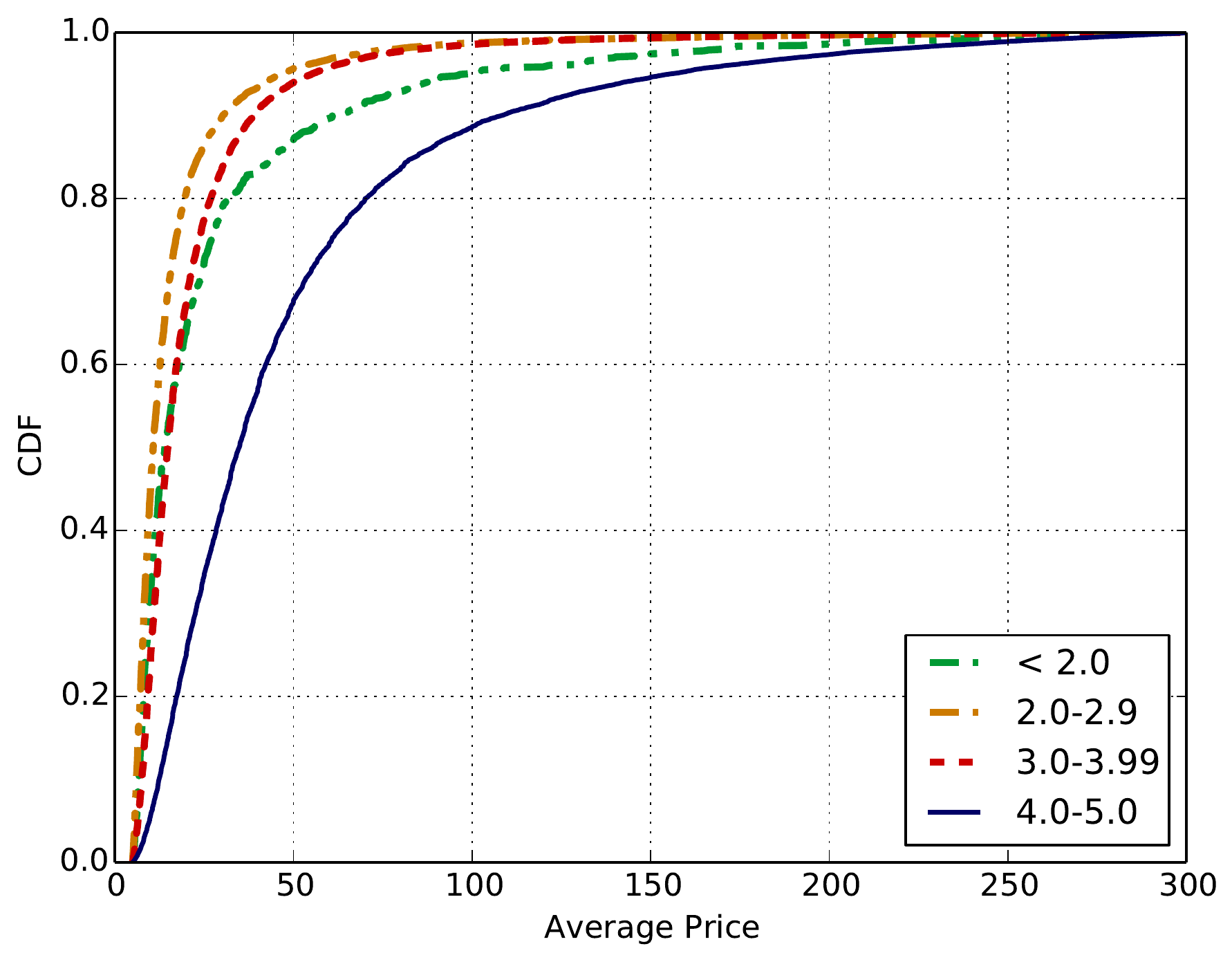}\label{fig:average_price_cdf_rating}}
    \caption{CDFs of the average price for the wines in our dataset.}
    \label{fig:average_price_cdf}
\end{figure*}

\descr{Ratings.} 
The ratings awarded to wines are generally favorable. The most common ratings are either 3.0 or 4.0 stars. All of the 21 wine-producing nations have a mean rating higher than 3.2, and no country has wines whose average rating exceeds 3.9. , Chilean (3.46), Brazilian (3.36), and Romanian (3.28) wines have the lowest average ratings. The highest average ratings belong to the USA (3.84), Germany (3.79), and France (3.78). The countries with the highest number of ratings per wine are Argentina (145.35), Chile (112.57), and Brazil (99.51). However, Italy (1.4M), France (1.1M) and the United States (1.0M) are the countries whose wines account for the largest number of total ratings. 
In Figure~\ref{fig:ratings_cdf}, we report the CDF of wine ratings divided by country (a) and of vintages by rating (b).

\descr{Prices.} As mentioned, we also collected the (average) price in pound sterling for each wine, when available. We find interesting differences across countries. The country with the most expensive wines is France (\textsterling 127.12 average price); this is inline with the fact that a disproportionate number (82\%) of the world's most expensive wines are produced in France \cite{winesearcher}.   
After France, we find the United States (\textsterling 67.54), followed by Portugal (\textsterling 64.69). %
Conversely, the three countries with the lowest mean prices are Croatia (\textsterling 15.94), South Africa (\textsterling 15.54), and Romania (\textsterling 11.62). 
In Figure~\ref{fig:average_price_cdf}, we report the CDFs of wine prices divided by country (a) and ratings (b).

\descr{User Reviews.} Finally, we find that wines from USA, France, and Italy receive the bulk of the written reviews -- 209.8k, 139.6k, and 132.1k, respectively (see Table \ref{tab:main_stats}), in addition to the highest number of replies to these review posts, with 127.7k, 148.7k, and 117.7k. Whereas, the highest levels of engagement, i.e. replies per comment, are for wines from Croatia (1.800), United Kingdom (1.689) and Israel (1.321). The levels of engagement for wines from USA, France and Italy are 0.609, 1.065, and 0.891. The mean engagement for all countries is 1.25 replies per post.

\begin{figure*}[t]
    \centering
    \subfigure[Country]{\includegraphics[width=0.375\textwidth]{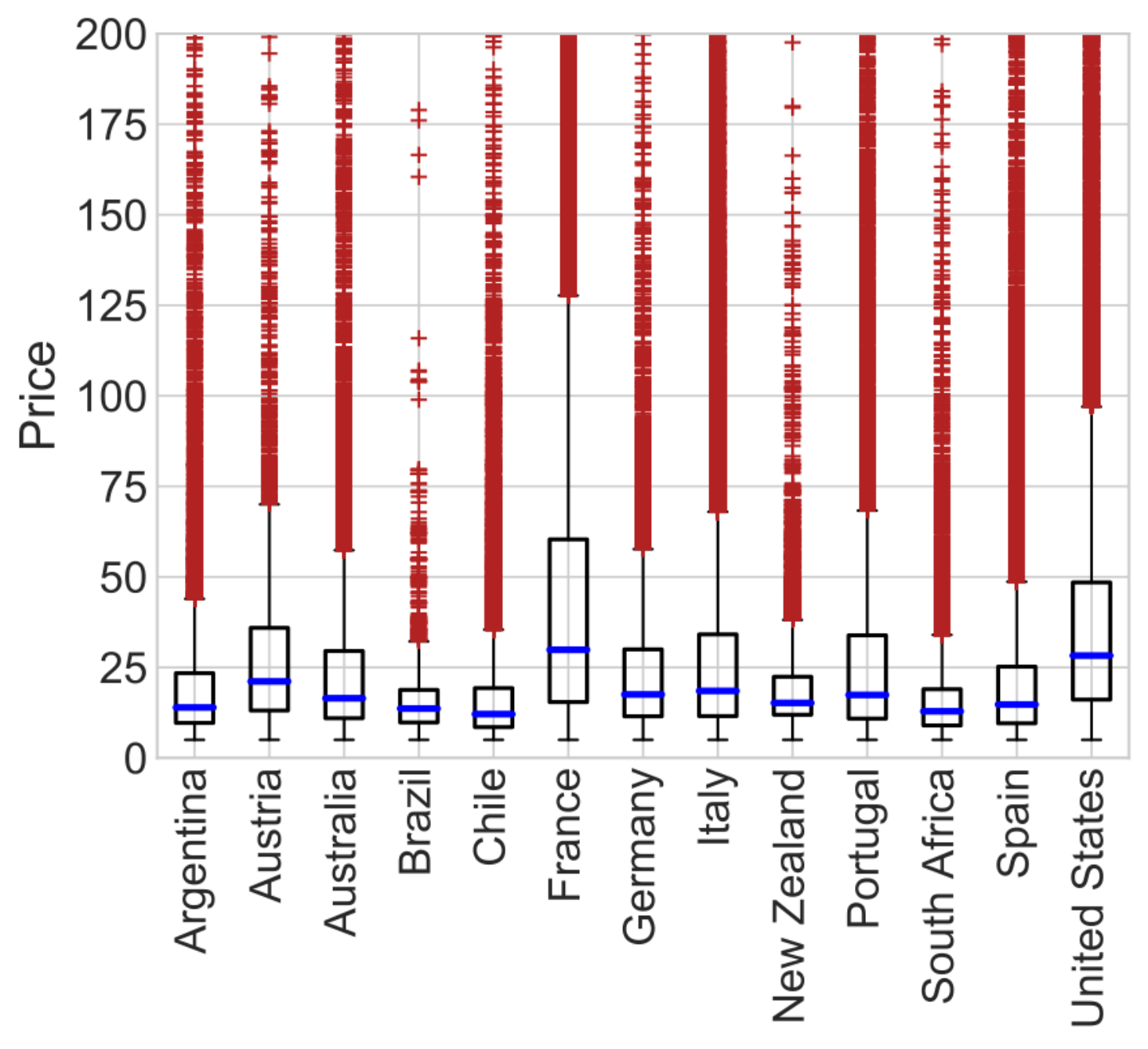}}
    ~~~
    \subfigure[Type]{\includegraphics[width=0.375\textwidth]{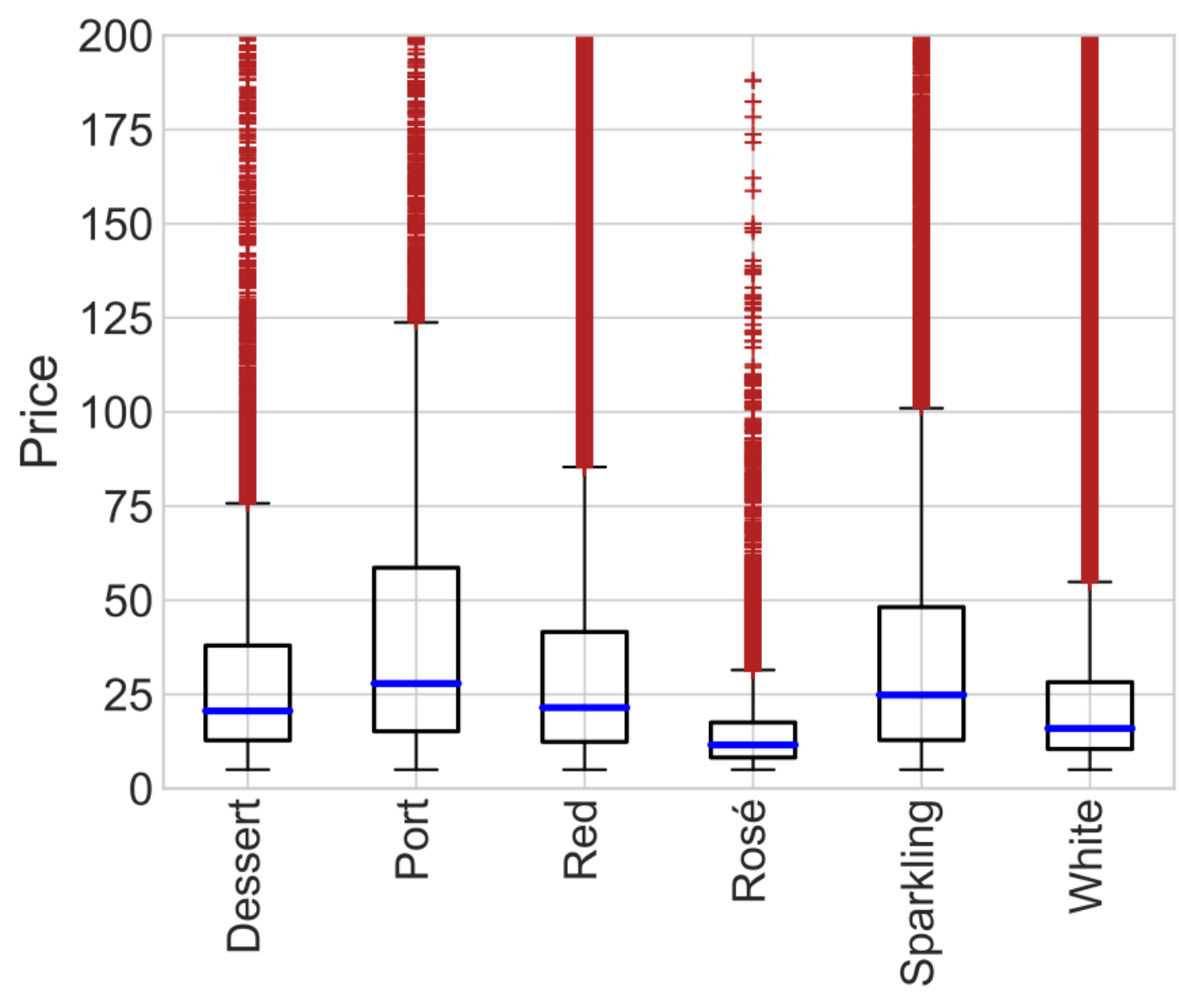}}
    \subfigure[Single Varietal Wines]{\includegraphics[width=0.375\textwidth]{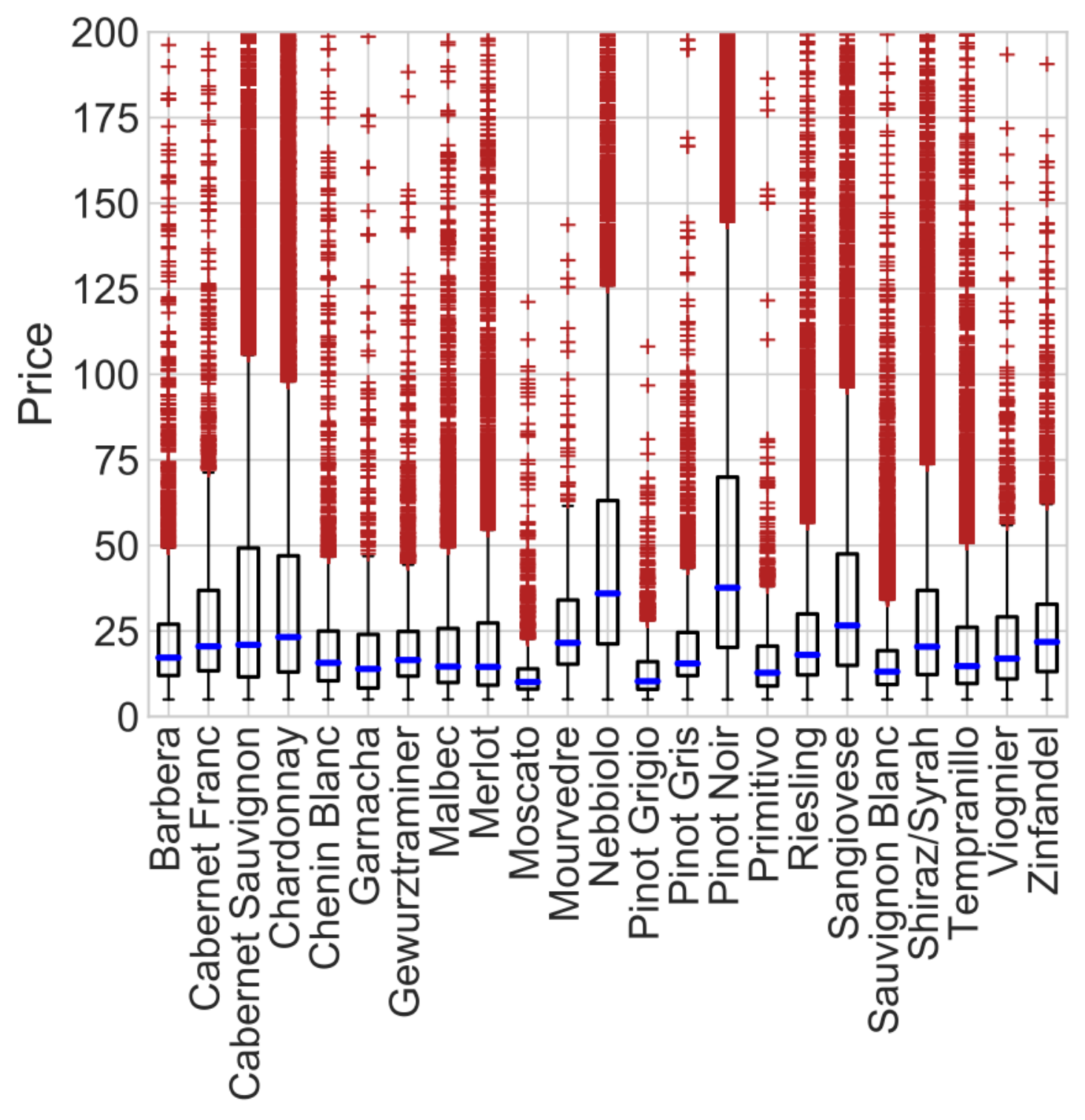}}
	~~~
    \subfigure[Blended Wines]{\includegraphics[width=0.375\textwidth]{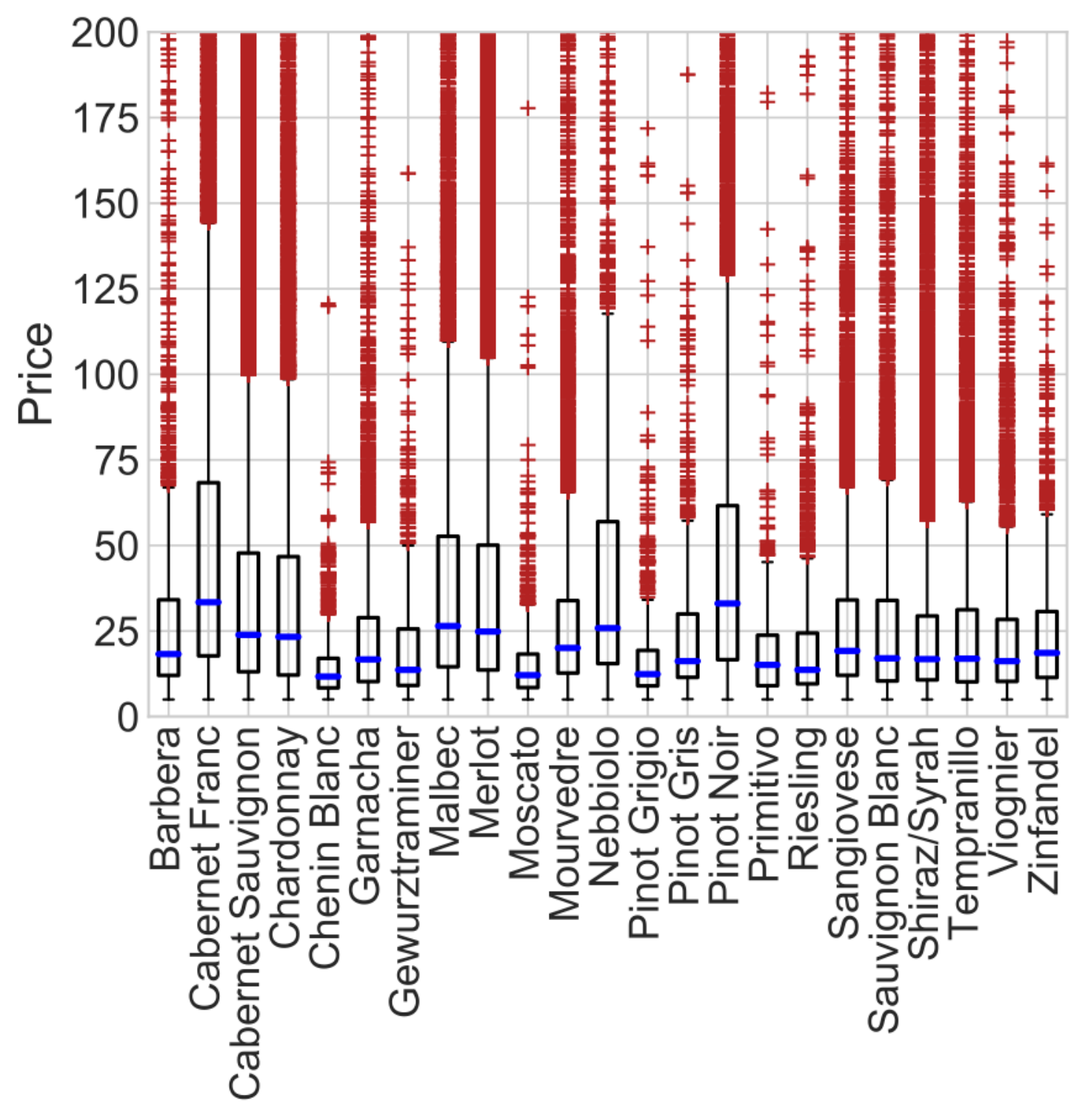}}
    \caption{Average price distributions in pound sterling, shown for wines with average prices in the interval [5,10$^6$). Plots are truncated at 200.}
    \label{fig:price_distributions}
        \vspace*{-0.1cm}
\end{figure*}

\section{Rating Analysis}\label{sec:ratings}

As mentioned, Vivino users assign ratings to wines  on a scale between 1.0 and 5.0 stars, with a 0.5 granularity.
We now analyze the Vivino wine ratings, investigating the relationship between them and prices, origins, types, style, and grapes. %

\subsection{Rating Trends}

\descr{Evolution over time.} Expert wine reviews have at times been criticized for a perceived inflation, i.e., that the preferences of wine suppliers for highly rated wines influences the reviews produced by experts~\cite{teeter2014}. (Naturally, the higher the rating of a wine, the greater the demand, which, according to market forces, also drives up the price.)
This trend has also somewhat manifested itself in the fast growing number of wines receiving a perfect score on the Robert Parker scale,
which went up from 17 in 2004 to 103 in 2013. Calls for new evaluation methods for wine quality have increased~\cite{snipes2014model}.
By contrast, we find that, regardless of type, country of origin, varietal, or blend, ``older'' wines, i.e. those produced between 1960 and 2000, are preferred to ``newer'' wines, i.e., those produced after 2000.
Across all wine variables: type, country, varietal, blend, there is a decline in average ratings between 2000 and 2010, but begin to increase again for 2016 vintages.

Grapes (both varietals and blends) and countries also show a similar ratings decline over the years. However, of the four most popular wine-producing countries: France, Italy, Spain and the USA, wines from the latter show the least decline (from 3.87 to 3.83 a decrease of 0.04 in the period between 2000 and 2010), whereas, Italian wines show a decline of 0.21, French wines experience a decline of 0.26, and Spanish wines fall in popularity by 0.17 stars. As we discuss later, %
this is partially due to biases of Vivino users, and in particular American users, for local wines. One reason for the general trend in falling ratings could be that users have a more positive perception of older wines. Another reason is that as the number of users on the social network has grown, the distribution of wine ratings may have changed to reflect a more diverse (and perhaps discerning) user base. 

\subsection{Wine Ratings and Prices}

We first examined the trend in prices across (a) country, (b) type, (c) varietals, and (d) blends as shown in Figure \ref{fig:price_distributions}. 
Price listings on Vivino are based on averages generated from a limited set of vendors and the price data for wines has a low frequency (see Section \ref{sec:datasets}).
Therefore, we cannot guarantee the reproducibility of these results, however, our analysis can provide an indication of the correlation between price and ratings.
We find that prices vary substantially across wine variables, in particular vintage, as expected older wines are more expensive than newer ones. Similarly, wines from specific countries: France in particular, but also the United States are more expensive than others. 

Next, we set out to test whether the average price of the wine correlates to high average ratings. We find no evidence to support a relationship that more expensive wines receive higher scores than less expensive wines. However, there does seem to be some relationship between price and ratings. Figure \ref{fig:average_price_cdf_rating} shows how the average price relates to ratings for wines in four ratings brackets. The highest average price is observed for wines rated between 2.0 and 3.0 stars, followed by wines in the 3.0 to 4.0 stars bracket. Interestingly, wines rated below 2.0 stars have a higher average price than those rated between 4.0 and 5.0 stars, whose average price is the lowest. This could be because more expensive wines do not meet the high expectations placed on them.

\begin{figure*}[t]
    \centering
    \subfigure[\#Followers]{\includegraphics[width=0.55\columnwidth]{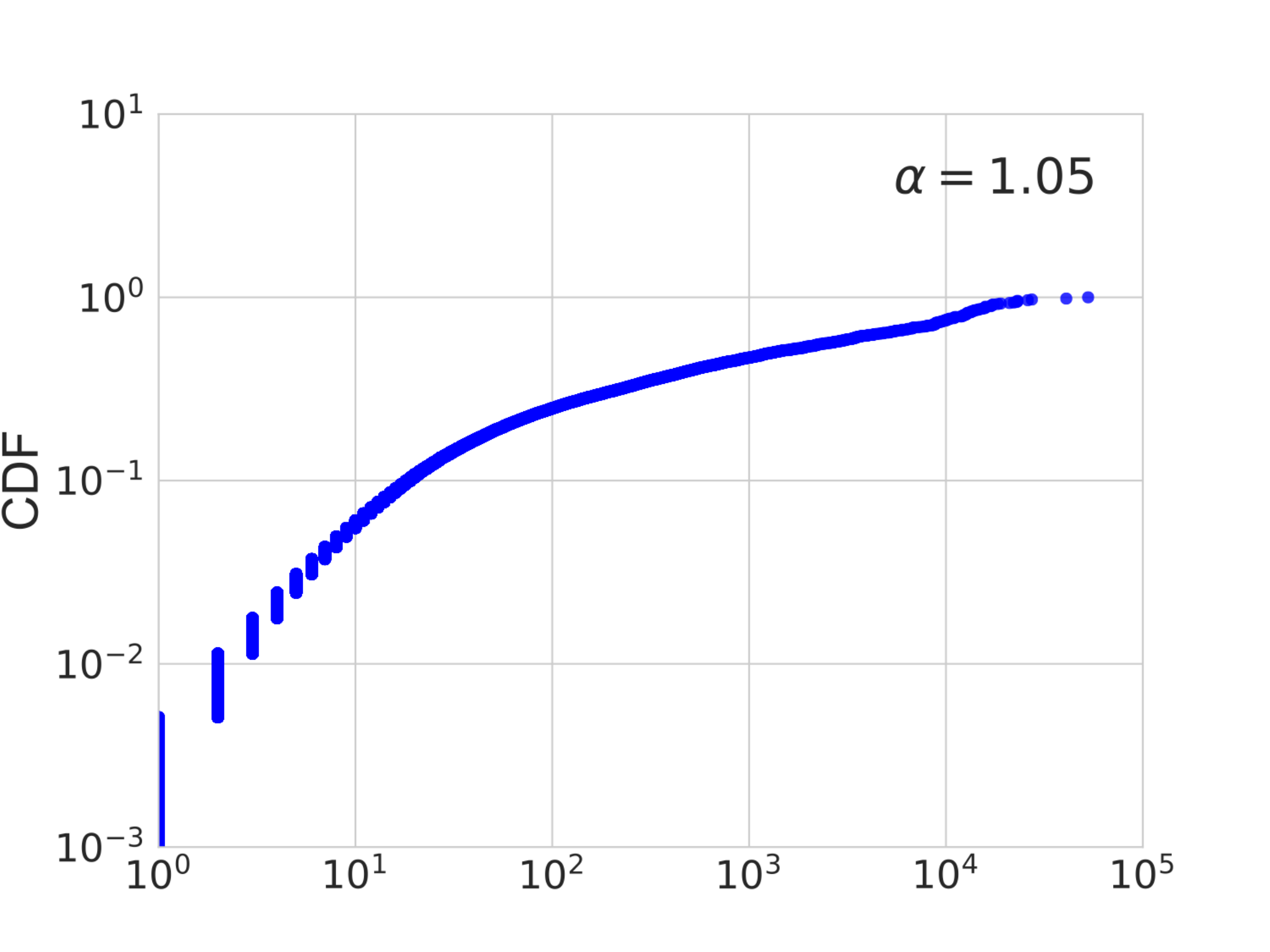}}
    ~~~
    \subfigure[\#Users Following]{\includegraphics[width=0.55\columnwidth]{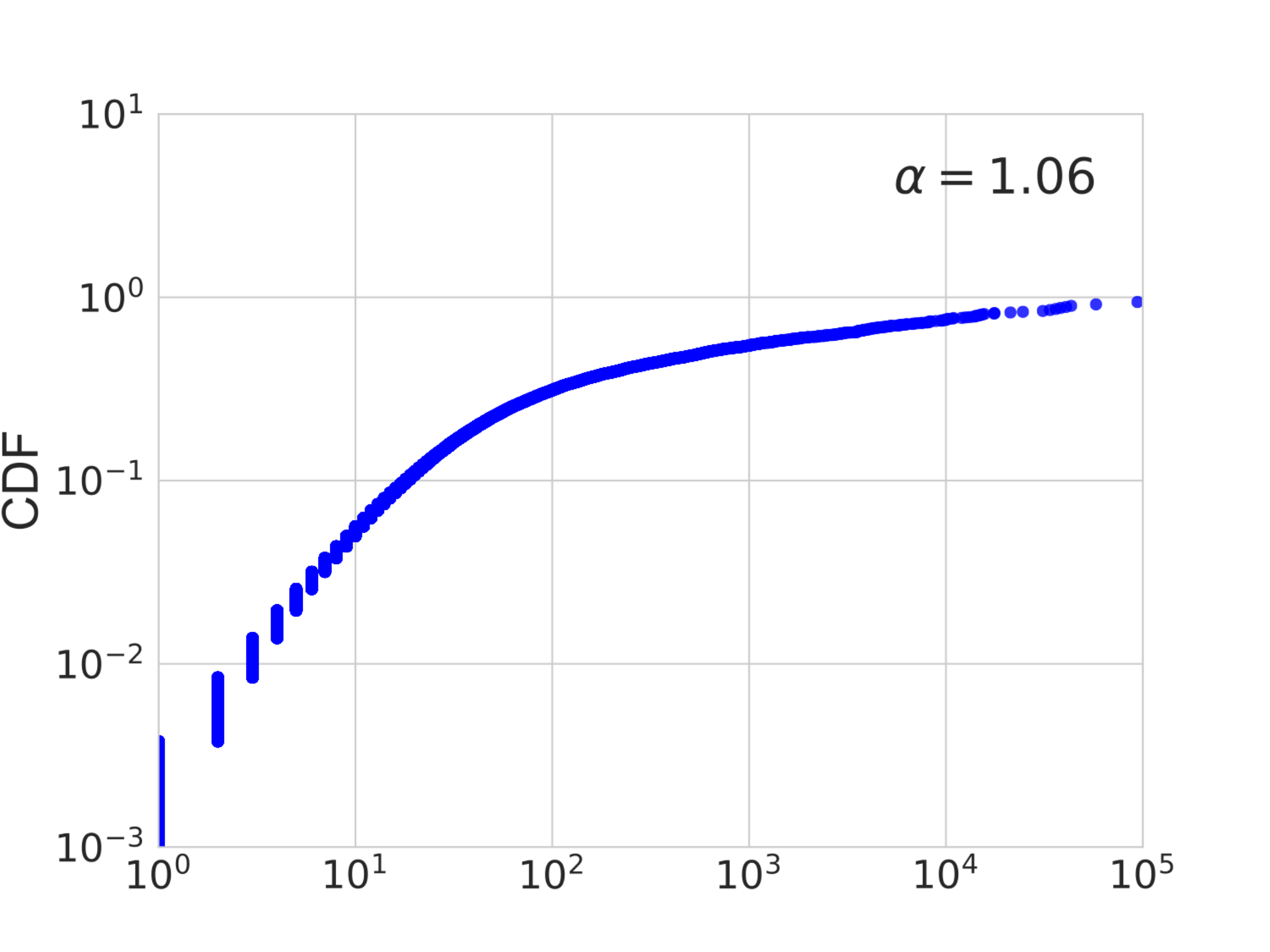}}
    ~~~
    \subfigure[\#Ratings]{\includegraphics[width=0.55\columnwidth]{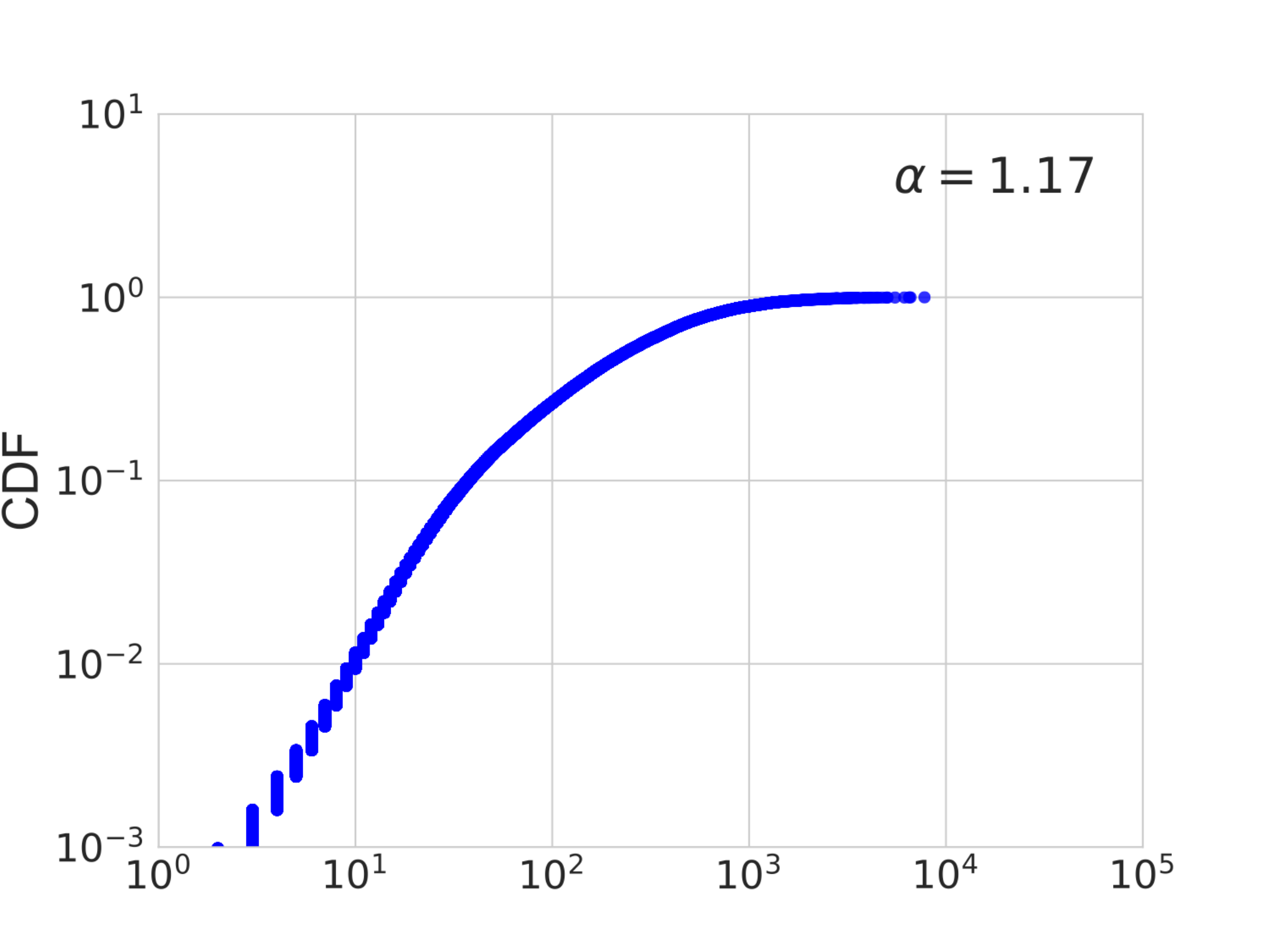}}
\caption{Distributions for the top 10k Vivino users from countries with the highest levels of wine consumption.}  \label{fig:distributions}
\end{figure*}

\begin{table*}[t]
    \centering
    \setlength{\tabcolsep}{8pt}
    \small
    \begin{tabular}{ r|lrr|lrr}
    \toprule
    {\bf \#} & {\bf Comment n-gram} & {\bf Freq.} & {\bf LR} & {\bf Reply n-gram} & {\bf Freq.} & {\bf LR}\\
    \midrule
    1 & ('full', 'bodied') & 1916 & 2455.13 & ('long', 'finish') & 93 & 141.959 \\ 
    2 & ('long', 'finish') & 1574 & 1976.26 & ('full', 'bodied')  & 88 & 136.86 \\ 
    3 & ('easy', 'drinking')  & 1244 & 1773.29 & ('pinot', 'noir')  & 51 & 97.2578 \\ 
    4 & ('well', 'balanced') & 1574 & 1602.5 & ('thanks', 'sharing') & 77 & 82.3297\\
    5 & ('pinot', 'noir') & 854 & 1584.19 & ('easy', 'drinking') & 49 & 74.0608\\
    6 & ('easy', 'drink') & 941 & 1198.65 & ('looking', 'forward') & 37 & 72.807\\
    7 & ('good', 'value')  & 1499 & 1139.07 & ('black', 'cherry') & 52 & 69.0606\\
    8 & ('green', 'apple') & 577 & 1126.39 & ('medium', 'body')  & 48 & 68.653\\
    9 & ('medium', 'bodied') & 788 & 922.062 & ('well', 'balanced') & 68 & 68.1572\\
    10 & ('fruit', 'forward') & 807 & 831.71 & ('dark', 'fruits') & 44 & 58.9347\\
    \bottomrule
    \end{tabular}
    \caption{Top 10 n-grams from wine reviews comments and replies ranked according to Likelihood Ratio (LR). }
    \label{tab:ngrams_table}
\end{table*}

\subsection{Use of Language on Vivino}
For our analysis of language employed by Vivino users, we examine the biographies and reviews of users. User biographies are tag-lines and short snippets of text, which appear on a profile page, they often state a user's motivation for joining the social network and occasionally provide a link to the individual's personal or professional websites. User wine reviews, which are typically accompanied by a quantitative score, vary from single word comments, expressing either approval or disapproval, to lengthier descriptive texts, which outline a user's experiences of a wine. These usually detail the taste and appearance of the beverage, and suggest dishes which complement the wine and from time to time describe occasions where the wine could be served.

\subsubsection{User Biographies}
Only 16k (11.6\%) of the 137.5k high-ranking users have biographies. 
This finding, coupled with the power-law distributions plotted in Figure~\ref{fig:distributions}, which show $\alpha$ = 1.05, $\alpha$ = 1.06, and $\alpha$ = 1.17 for followers, users following, and ratings respectively,
indicates that wine ratings, and not social connections, are the main motivation behind users' interactions on the platform. 

683 biographies contain the word \textit{sommelier}; 434 biographies contain the abbreviations \textit{WSET}, \textit{N2}, \textit{N3}, \textit{N4} or \textit{N5} (names of professional wine tasting qualifications); 120 contain the word \textit{expert}; and 135 contain the word \textit{professional}. Conversely, 300 users describe themselves as \textit{amateur}, 159 as \textit{learning}, and 9 think of themselves as a \textit{learner}. Interestingly, significantly more Vivino biographies contain keywords implying that users are wine experts, than those which contain keywords implying they are novices. However, the sample size is very small, and therefore few findings can be confidently drawn from these data. Furthermore, 274 biographies contain a web address, of these, 114 are email addresses. Thus, it would be fair to assume that these users are looking to forge professional connections, rather than social ones.

We also examine how user biographies vary across region, ranking and number of ratings contributed. 
Biography unigram and bigram frequency distributions do not vary significantly by user ranking. However, they do when users are grouped by the number of ratings contributed. For users with fewer ratings, their bios typically have a high frequency of the words \textit{learning}, \textit{student}, \textit{amateur} and \textit{enthusiast}. On the other hand, for users who have contributed more than 1,000 wine ratings on Vivino, the frequency for \textit{enthusiast} is significantly lower. These users are much more likely to describe themselves as \textit{certified}, having a WSET qualification and provide their email address in their biography.

\subsubsection{Wine Reviews}
We analyze 
reviews published between the dates 8 September 2012 and 27 March 2017. These comments, which we call reviews in this section, are also associated with replies or feedback posts. There is a one-to-many relationship between reviews and replies. 
The contents of replies are not of as much interest to us as the content of reviews: typically, replies express agreement or gratitude as shown in Table \ref{tab:ngrams_table}. In fact, the two most common bigram collocations for replies are  (\textit{dear}, \textit{thank}), with a likelihood ratio of 77.21, and (\textit{Thanks}, \textit{sharing}),  with a likelihood ratio of 69.87. Recall that a collocation is a sequence of tokens that appears with a high probability in a text, and can be determined, e.g., using the Natural Language Toolkit (NLTK) Python package \cite{BirdKleinLoper09}.
We use the likelihood ratio association metric to determine bigram collocations.
Mean word count is 64.05, the standard deviation is 37.33. Although the written reviews provided by Vivino users are short in length, they contain a wide range of vocabulary. In particular, the words employed by users to describe wines are taken from a lexicon commonly employed by professional wine critics \cite{puckette2012}. 
The most frequently occurring descriptors are: \textit{good}, \textit{nice}, \textit{great}, \textit{smooth}, \textit{dry}, \textit{fruity}, \textit{light}, \textit{red}, \textit{sweet}, \textit{well}. The most common unigrams include \textit{tannins} and \textit{acidity} -- which are also among the most common words used in professional reviews \cite{puckette2012}.

\section{Wine Prediction Models}\label{sec:prediction}

We then set out to build two predictive models: (1) a regression model for predicting the average rating of a wine; and (2) a classification model for predicting the taste profile or reviews history of a Vivino user. For the former, we evaluate the performance of both a Decision Tree (DT) regressor and a Multilayer Perceptron Neural Network (MLP). For the classification task of predicting user taste profiles, we use Random Forests (RF). 

\begin{table}[t]
    \centering
    \setlength{\tabcolsep}{3.75pt}
    \small
    \begin{tabular}{ l  r  r  r  r }
    \toprule
    {\bf Regression Model} & {\bf Train $R^2$} & {\bf Test $R^2$} & {\bf Train MSE} & {\bf Test MSE}\\
    \midrule
     MLP & 0.645 & 0.640 & 0.0459 & 0.0469\\
     DT & 0.701 & 0.608 & 0.0383 & 0.0520\\
    \bottomrule
    \end{tabular}
    \caption{$R^2$ Scores and mean squared error (MSE) for wine ratings prediction models for a 75\% train set and 25\% test set. }
    \label{tab:regression_scores}
\vspace*{-0.1cm}
\end{table}

\subsection{Predicting Wine Ratings}

First, we develop feature engine for the model as part of the pre-processing stage. The task of constructing a good feature representation is essential in order to train supervised learning models effectively. 
As we aim to implement a generalized model, with capabilities of predicting the average rating of any given wine, we discount the following wine and winery specific columns of the wine dataset: wine name and winery name. We find that even though there are recurring noun phrases in the wine names, these noun phrases either referred to the grapes, which constituted to the wine, or the vintage or the wine's type -- data which is already present in other columns of the of the wine dataset. For this reason we decide not to include names in the feature representation. Also excluded from the feature representation were wines with less than 75 ratings. One hot encoding is used to represent the categorical data: region, country, wine type (e.g. red, white, sparking, port), regional style, food pairings and grapes. The outcome of the pre-processing stage was a feature representation with 4.4k dimensions.  Univariate feature selection was then employed to reduce the dimensions to 1k. The combined train and test sets accounted for 124,397 wines. A train-test ratio of 75\% : 25\% was selected for both the MLP and DT regressor models.  

Although the DT outperforms the MLP approach on the train set, the MLP shows the best generalization on the test set. The accuracy metric chosen to compare the two models is the coefficient of determination. This regression model error metric is commonly denoted as \begin{math}{R^2}\end{math}, where~\cite{nagelkerke1991note}:\vspace*{-0.15cm}
$$R^{2}=1-\sum_{i}(y_{i}-\hat{y_i})^{2}/\sum_{i}(y_{i}-\bar{y})^{^{2}}\vspace*{-0.15cm}$$
\begin{math}{R^2}\end{math} for both methods is shown in Table \ref{tab:regression_scores}. The DT model achieves an accuracy of 0.701 on the train set, whereas the MLP achieves a lower score of 0.640. For the test set, the accuracy of the DT model decreases by 13.3\%, whereas the MLP model only decreases by 0.78\%. This is most likely because the one hot coded representation of the wine features allows the MLP to generalize to unseen data more effectively. 

The maximum value for \begin{math}{R^2}\end{math} is 1.0. Negative values can also be produced for poorly performing models. 

\begin{math}{R^2}\end{math} for both methods is shown in Table \ref{tab:regression_scores}. 
The DT model achieves an accuracy of 70\% on the train set, whereas the MLP achieves a lower accuracy score of 64\%. When the models are performed on the test set, the accuracy of the DT model decreases by 15.4\%, whereas the MLP model only decreases by 0.744\%. This is most likely because the one-hot coded representation of the wine features allows the MLP to generalize to unseen data more effectively.

Table \ref{tab:feature_importance} shows the Gini index values of the ten most important features used to train the Decision Tree (DT) model. The feature with the highest Gini index is \textit{average price}, which is far greater than that of \textit{number ratings}, i.e., the second most important feature. The other eight features represent one hot-encoded features -- one grape, one food pairing, and the remaining five are related to wine regional styles. The least important features are \textit{region: Dao} (998th), \textit{year: 1858} (999th), and \textit{region: Knights Valley} (1000th).

\begin{table}[t]
    \centering
    \setlength{\tabcolsep}{3pt}
    \small
    \begin{tabular}{ r|lr}
    \toprule
    {\bf \#} & {\bf Decision Tree (DT) Feature} & {\bf Gini}\\
    \midrule
    1 & average price 						& 0.82400  	\\
    2 & number ratings						& 0.02160	 	\\ 
    3 & food pairing: poultry 				& 0.00954	\\ 
    4 & 	grape: Zinfandel						& 0.00678 	\\
    5 & regional style: Italian Vino Nobile Di Montepulciano	& 0.00588 	\\
    6 & regional style: Austrian Pinot Gris 	& 0.00502	\\
    7 & regional style: New Zealand Chardonnay &	0.00420	\\
    8 & regional style: Italian Ripasso 		  & 	0.00392	\\
    9 & regional style: Israeli Syrah 		  & 	0.00320	\\
    10 &	 regional style: Spanish Montsant Red  & 0.00314 \\
    11 &	 regional style: South African Malbec  & 0.00314	\\
    12 &	 food pairing: Cured meat 			   & 0.00286	 \\
	13 &	 regional style: South African Cabernet Franc	& 0.00249 \\
 	14 & grape: Cabernet Franc 							& 0.00232 \\
	15 &	 regional style: Northern Italy Muller Thurgau	& 0.00226 \\
	16 &	 regional style: Australian Viognier 			& 0.00222 \\
	17 &	 regional style: Greek Amyndeon Red				& 0.00214 \\
	18 &	 country: South Africa							& 0.00213 \\
	19 &	 grape: Mourvedre								& 0.00207 \\
	20 &	 regional style: South African Merlot			& 0.00207 \\
	
    \bottomrule
    \end{tabular}
    \caption{Gini importance for Decision Tree (DT) features used to train wine ratings predictors, in descending order of importance.  
    }
    \label{tab:feature_importance}
    \vspace*{-0.2cm}
\end{table}

\begin{figure*}[t]
   \centering
   \begin{minipage}[t]{.49\textwidth}
	\centering
	\includegraphics[width=0.8\textwidth]{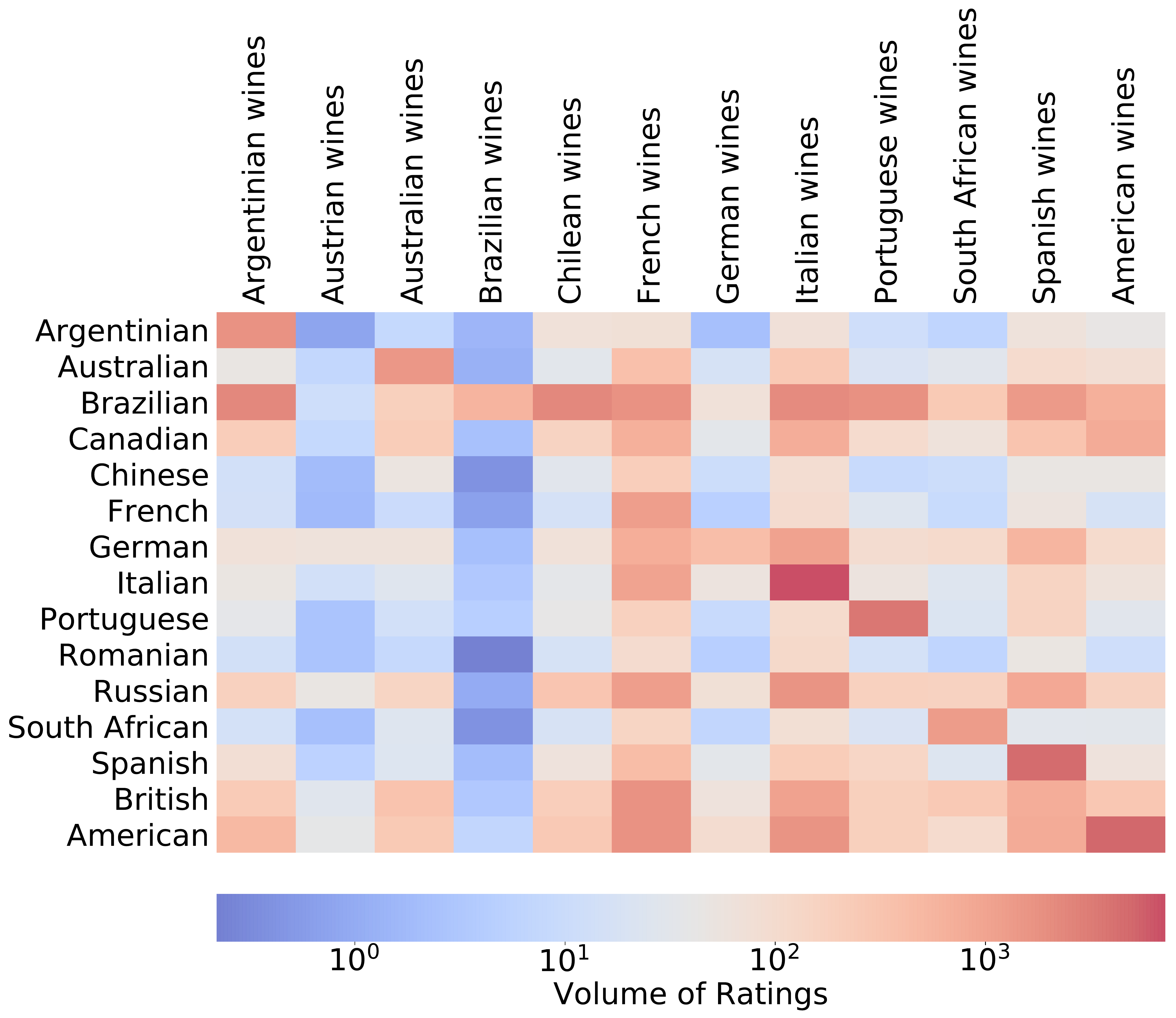}
	\caption{The total volume of ratings of wines from users from 15 countries with the highest levels of wine consumption in 2015 \protect\cite{oiv2016}.}
	\label{fig:most_commonly_rated}
\end{minipage}
~~
   \begin{minipage}[t]{.49\textwidth}
\centering
	\includegraphics[width=0.8\textwidth]{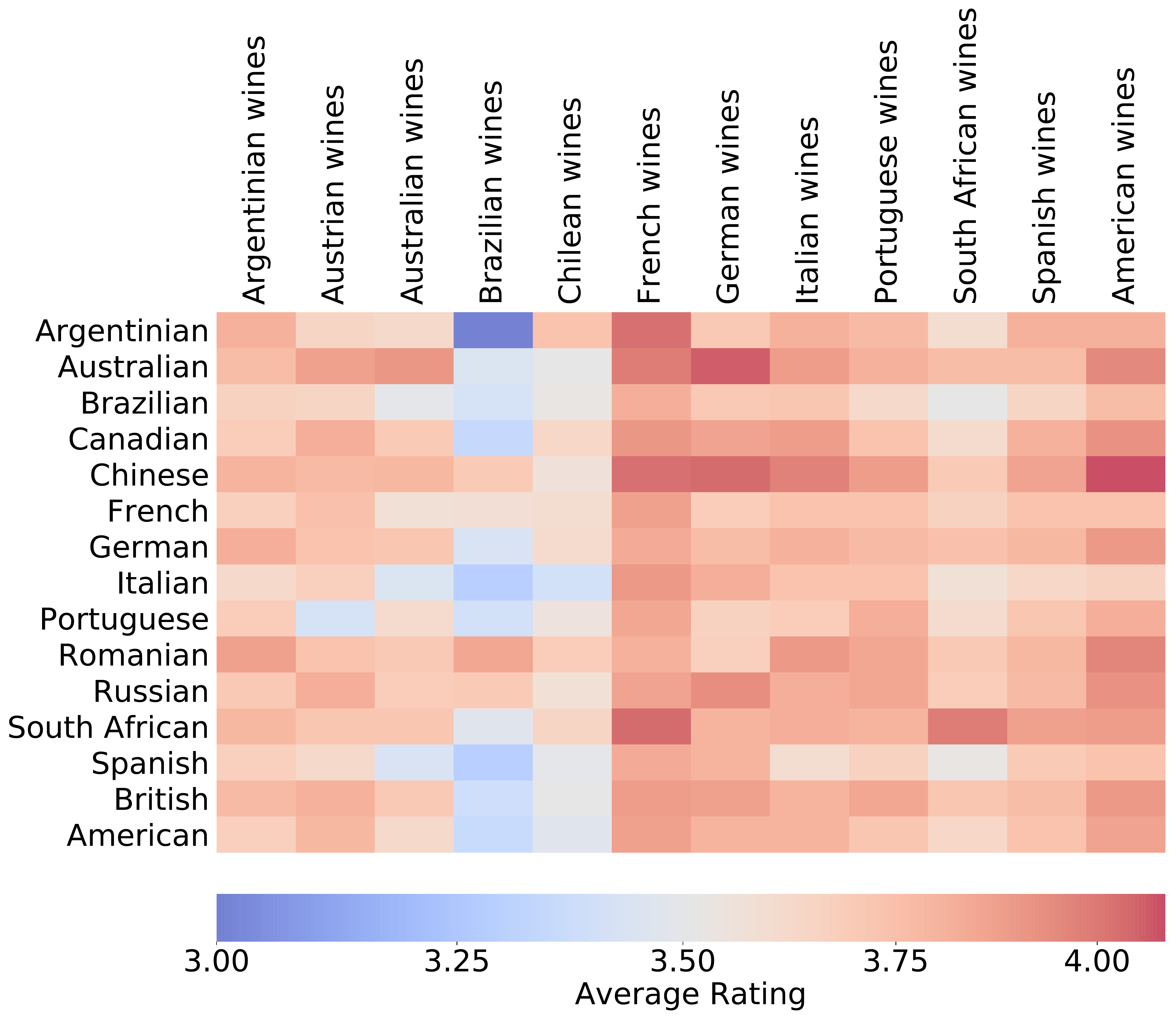}
	\caption{The average rating of wines from users from 15 countries with the highest levels of wine consumption in 2015 \protect\cite{oiv2016}.}
	\label{fig:most_highly_rated}
\end{minipage}
\end{figure*}

\begin{figure*}[t]
\centering
   \begin{minipage}[t]{.49\textwidth}
\centering
	\includegraphics[width=0.9\textwidth]{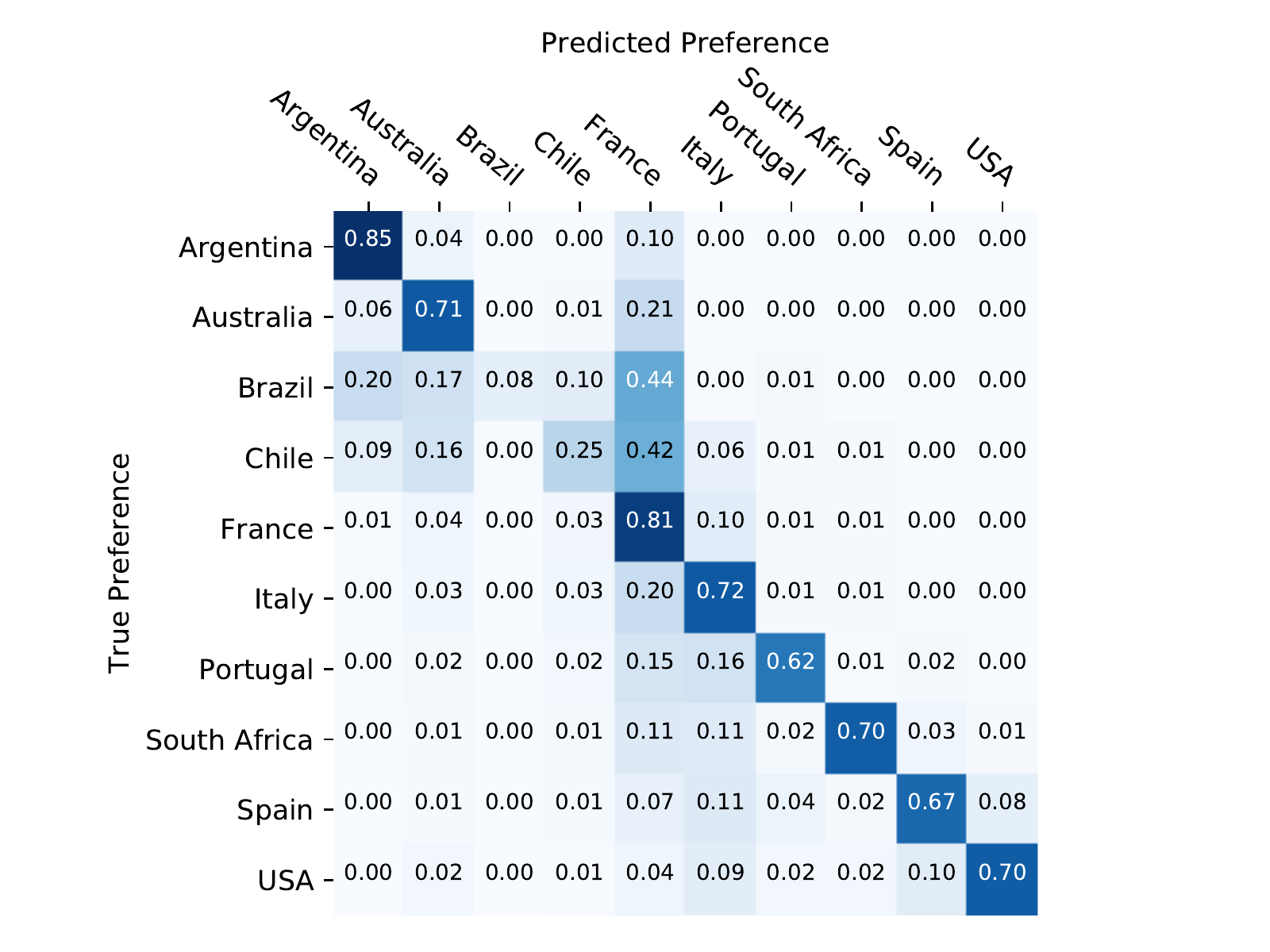}
	\caption{Confusion matrix for user preferences classifier based on the 10 countries with the most ratings in the wine dataset.}
	\label{fig:confusion_matrix_wine_data}
\end{minipage}
~~
   \begin{minipage}[t]{.49\textwidth}
\centering
	\includegraphics[width=0.9\textwidth]{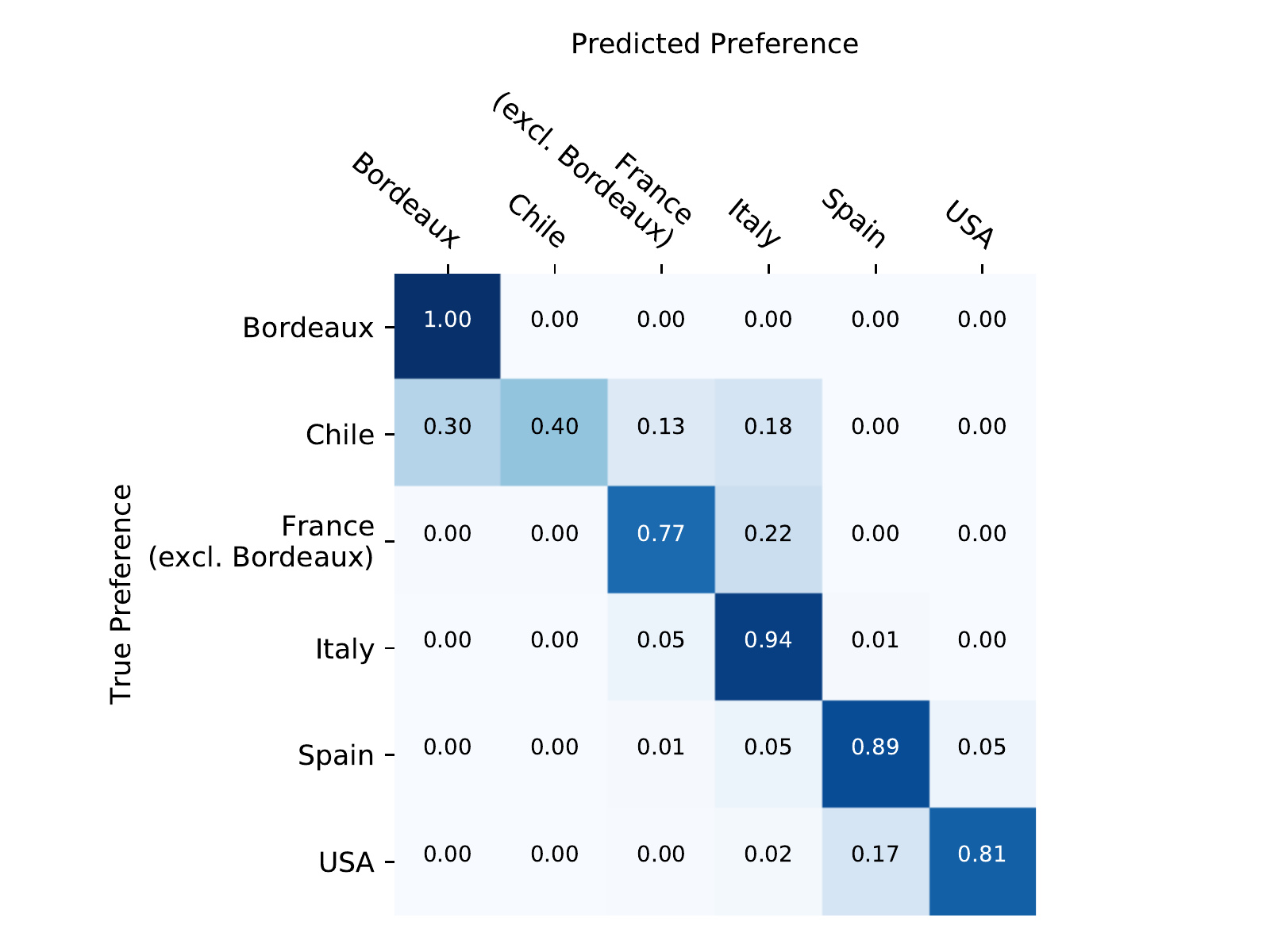}
	\caption{Confusion matrix for user preferences classifier based on the countries/regions with the most ratings in the user taste profiles.}
	\label{fig:confusion_matrix_user_data}
	\end{minipage}
\end{figure*}

\subsection{Predicting Wine Preferences}

Next, we attempt to use the data analysis in order to create a wine preferences classifier. Before we begin to design this model, we first extend our wine ratings analysis, with a particular focus on users. From this analysis, we learn that users show a preference for rating wines which originate in their home countries. Figure \ref{fig:most_commonly_rated} shows that for the majority of countries, Vivino users are most likely to rate a local wine. For the countries for which this is not the case, users either have a strong preference for rating wines that are produced in nearby countries or European wines. The most commonly rated wines are not necessarily the most highly rated wines, as shown in Figure \ref{fig:most_highly_rated}. Brazilian wines receive the lowest volume of ratings from all Vivino users, apart from Brazilian users themselves. As well as receiving the lowest volume of ratings, Brazilian wines also receive the greatest proportion of ratings below 3.5. Aside from the USA, New World wines (from Brazil, Chile, Argentina, Australia and South Africa) receive lower average ratings that Old World wines. 

Preferences are acquired from the user taste profile: a history of average ratings and number of ratings for each regional style a Vivino user has reviewed.
Each example in the user set is labelled as either a user with a preference for wines from Argentina, Australia, Brazil, Chile, France, Italy, Portugal, South Africa, Spain or USA. These ten countries are chosen as their wines accounted for the majority of the taste profile samples. The Random Forests algorithm is used to classify user preferences. The mean accuracy of the model is 79.8\%. 

We develop two versions of the classifier: 
\begin{compactenum}
\item A model differentiating between the preferences for wines from the 10 most commonly rated countries (from the wine dataset): Argentina, Australia, Brazil, Chile, France, Italy, Portugal, South Africa, Spain, and United States. The accuracy for this model is 71.8\%. 

\item A second model based on the country of origin of wines most frequently reported on in the taste profiles in the user dataset. These countries and regions are Bordeaux, Chile, France (excluding Bordeaux), Italy, Spain, and United States. This model achieves a higher accuracy, namely, 79.8\%. The confusion matrices for these two models are shown in Figure \ref{fig:confusion_matrix_wine_data} and Figure \ref{fig:confusion_matrix_user_data}. 

\end{compactenum}

We also report the confusion matrix for the classifier in Figure \ref{fig:confusion_matrix_user_data}. The confusion matrix shows that with the exception of wines from France (excluding Bordeaux), which are most often confused for Bordeaux, all other wine preferences are most often confused for wines from France (excluding Bordeaux).

\section{Conclusion}\label{sec:conclusion}

In this paper, we examined how user biases, regional factors and wine characteristics affect wine ratings on Vivino, a social network application for wine enthusiasts. 
Our analysis found that the biggest indicator of how a user will rate a wine is its vintage and region, i.e., there is a strong preference for French wine styles, and users also have a more favorable opinion of home-grown wines. Furthermore, our work shows that more expensive wines do not necessarily receive higher ratings. This is in contrast to the bias researchers have found exists among professional critics~\cite{teeter2014}. Language analysis also shows that Vivino users generate relatively high quality user reviews, showing that wine ratings and reviews produced by amateur wine enthusiasts can be quite useful.

We also aimed to explore how wine characteristics, natural language analysis of wine reviews and regional analysis of wineries on Vivino could be integrated to develop a practical model for predicting wine reviews. We developed two predictions based on average wine ratings and users' ratings histories (which wines users rate and how they rate them). The models were evaluated using unseen examples from the dataset to gauge their efficacy. The results demonstrate that there is consistency across the ratings given by Vivino users, thus, spam and/or troll content does not affect the credibility of ratings on the social network. Our model also shows that wine ratings are not random, but the ratings assigned to wines by users based on informed and considered decisions. Overall, we believe that the analysis of factors that influence wine ratings and the development of models for predicting wine reviews are useful contributions to the understanding of how specialist social media platforms influence and shape our eating and drinking habits, and how we can minimize the subjectivity of online food and drinking ratings. 

On the other hand, we acknowledge that our models do have some limitations.
For instance, the user data collected was not of the same granularity as the wine data, and this may be one reason for the poorer performance of our preference classifier compared to the wine ratings regression model. The models produced were only tested on Vivino datasets, thus, as part of future work, we plan to evaluate them on user and wine reviews data from other sources.

\small
\bibliographystyle{abbrv}

\begin{thebibliography}{10}

\bibitem{abbar2015you}
S.~Abbar, Y.~Mejova, and I.~Weber.
\newblock You tweet what you eat: Studying food consumption through twitter.
\newblock In {\em CHI}, 2015.

\bibitem{atkins2016food}
P.~Atkins, I.~Bowler, et~al.
\newblock {\em Food in society: economy, culture, geography}.
\newblock Routledge, 2016.

\bibitem{BirdKleinLoper09}
S.~Bird, E.~Klein, and E.~Loper.
\newblock {\em {Natural Language Processing with Python}}.
\newblock O'Reilly Media, 2009.

\bibitem{chorley2016pub}
M.~Chorley, L.~Rossi, G.~Tyson, and M.~Williams.
\newblock Pub crawling at scale: tapping untappd to explore social drinking.
\newblock In {\em ICWSM}, 2016.

\bibitem{eufic}
{European Food Information Council}.
\newblock {Why we eat what we eat: Social and economic determinants of food
  choice}.
\newblock \url{http://bit.ly/2Dx8RpA}, 2018.

\bibitem{faloutsos1999power}
M.~Faloutsos, P.~Faloutsos, and C.~Faloutsos.
\newblock {On power-law relationships of the Internet topology}.
\newblock In {\em ACM CCR}, 1999.

\bibitem{ghewaripredicting}
R.~Ghewari and S.~Raiyani.
\newblock Predicting cuisine from ingredients.
\newblock {\em University of California San Diego}, 2015.

\bibitem{guidry2015mcdonaldsfail}
J.~D. Guidry, M.~Messner, Y.~Jin, and V.~Medina-Messner.
\newblock From \#mcdonaldsfail to \#dominossucks: An analysis of instagram
  images about the 10 largest fast food companies.
\newblock {\em Corporate Communications: An International Journal}, 20(3),
  2015.

\bibitem{higgs2016social}
S.~Higgs and J.~Thomas.
\newblock Social influences on eating.
\newblock {\em Current Opinion in Behavioral Sciences}, 9, 2016.

\bibitem{mejova2016fetishizing}
Y.~Mejova, S.~Abbar, and H.~Haddadi.
\newblock {Fetishizing Food in Digital Age: \#Foodporn Around the World}.
\newblock In {\em ICWSM}, 2016.

\bibitem{mejova2015foodporn}
Y.~Mejova, H.~Haddadi, A.~Noulas, and I.~Weber.
\newblock {\#FoodPorn: Obesity patterns in culinary interactions}.
\newblock In {\em ICDH}, 2015.

\bibitem{nagelkerke1991note}
N.~J. Nagelkerke.
\newblock A note on a general definition of the coefficient of determination.
\newblock {\em Biometrika}, 78(3):691--692, 1991.

\bibitem{oiv2016}
{OIV}.
\newblock State of the vitiviniculture world market.
\newblock
  \url{http://www.oiv.int/public/medias/4710/oiv-noteconjmars2016-en.pdf},
  2016.

\bibitem{oiv2017}
{OIV}.
\newblock Global economic vitiviniculture data.
\newblock
  \url{http://www.oiv.int/public/medias/5681/en-communiqu-depresse-octobre-2017.pdf},
  2017.

\bibitem{parker2002parker}
R.~M. Parker and P.-A. Rovani.
\newblock {\em Parker's wine buyer's guide}.
\newblock Simon and Schuster, 2002.

\bibitem{puckette2012}
M.~Puckette.
\newblock 40 wine descriptions and what they really mean.
\newblock http://winefolly.com/tutorial/40-wine-descriptions, April 2012.

\bibitem{sajadmanesh2016kissing}
S.~Sajadmanesh, S.~Jafarzadeh, S.~A. Ossia, H.~R. Rabiee, H.~Haddadi,
  Y.~Mejova, M.~Musolesi, E.~De~Cristofaro, and G.~Stringhini.
\newblock {Kissing Cuisines: Exploring Worldwide Culinary Habits on the Web}.
\newblock In {\em WWW Web Science Track}, 2017.

\bibitem{winesearcher}
W.~Searcher.
\newblock {The World's Top 50 Most Expensive Wines}.
\newblock \url{http://www.wine-searcher.com/most-expensive-wines}, 2017.

\bibitem{sharma2015measuring}
S.~S. Sharma and M.~De~Choudhury.
\newblock Measuring and characterizing nutritional information of food and
  ingestion content in instagram.
\newblock In {\em WWW}, 2015.

\bibitem{silva2014you}
T.~H. Silva, P.~O. de~Melo, J.~Almeida, M.~Musolesi, and A.~Loureiro.
\newblock {You are what you eat (and drink): Identifying cultural boundaries by
  analyzing food \& drink habits in foursquare}.
\newblock In {\em ICWSM}, 2014.

\bibitem{snipes2014model}
M.~Snipes and D.~C. Taylor.
\newblock Model selection and akaike information criteria: An example from wine
  ratings and prices.
\newblock {\em Wine Economics and Policy}, 3(1), 2014.

\bibitem{teeter2014}
A.~Teeter.
\newblock How a ratings system designed to help consumers is tearing the \$40
  billion dollar wine industry apart., Nov. 2014.

\bibitem{tiu2001food}
L.~Tiu~Wright, C.~Nancarrow, and P.~M. Kwok.
\newblock Food taste preferences and cultural influences on consumption.
\newblock {\em British Food Journal}, 103(5):348--357, 2001.

\bibitem{trevisan2011wine}
M.~Trevisan.
\newblock Wine and society.
\newblock {\em Wine Studies}, 1(1), 2011.

\end{thebibliography}

%

\end{document}